\documentclass[reprint,superscriptaddress,nofootinbib,nobibnotes,amsmath,amssymb,aps,prd,longbibliography,twocolumn, noeprint]{revtex4-2}

\usepackage{float}
\usepackage[percent]{overpic}
\usepackage{graphicx, comment}
\usepackage{dcolumn}
\usepackage{bm}
\usepackage{xcolor}
\usepackage{hyperref}
\usepackage[mathlines]{lineno}
\usepackage[normalem]{ulem}
\usepackage{amsfonts}
\usepackage{url}

\usepackage{xurl}
\usepackage{makecell}
\usepackage{hhline}

\usepackage[inkscapelatex=false]{svg}

\usepackage{array}
\usepackage[normalem]{ulem}

\begin{document}

\newcommand{\red}[1]{\textcolor{red}{#1}}


\newcommand{\nuecc}{$\nu_e$CC}
\newcommand{\numuccpio}{$\nu_\mu$CC\ $\pi^0$}
\newcommand{\ncpio}{NC\ $\pi^0$}
\newcommand{\ncdelta}{NC\ $\Delta $}
\newcommand{\ncdeltarad}{NC\ $\Delta\rightarrow N \gamma$}
\newcommand{\onegOnep}{$1 \gamma 1 p$}
\newcommand{\onegZerop}{$1 \gamma 0 p$}
\newcommand{\onegNp}{$1 \gamma N p$}
\newcommand{\epem}{$e^+ e^-$}

\newcommand{\italicheading}[1]{
  \vspace{1em}
  \noindent\textit{#1}
}

\setlength{\parskip}{0pt}

\preprint{APS/123-QED}

\title{Enhanced search for neutral current $\Delta$ radiative single-photon production in MicroBooNE}

\newcommand{\ANL}{Argonne National Laboratory (ANL), Lemont, IL, 60439, USA}
\newcommand{\Bern}{Universit{\"a}t Bern, Bern CH-3012, Switzerland}
\newcommand{\BNL}{Brookhaven National Laboratory (BNL), Upton, NY, 11973, USA}
\newcommand{\UCSB}{University of California, Santa Barbara, CA, 93106, USA}
\newcommand{\Cambridge}{University of Cambridge, Cambridge CB3 0HE, United Kingdom}
\newcommand{\CIEMAT}{Centro de Investigaciones Energ\'{e}ticas, Medioambientales y Tecnol\'{o}gicas (CIEMAT), Madrid E-28040, Spain}
\newcommand{\Chicago}{University of Chicago, Chicago, IL, 60637, USA}
\newcommand{\Cincinnati}{University of Cincinnati, Cincinnati, OH, 45221, USA}
\newcommand{\CSU}{Colorado State University, Fort Collins, CO, 80523, USA}
\newcommand{\Columbia}{Columbia University, New York, NY, 10027, USA}
\newcommand{\Edinburgh}{University of Edinburgh, Edinburgh EH9 3FD, United Kingdom}
\newcommand{\FNAL}{Fermi National Accelerator Laboratory (FNAL), Batavia, IL 60510, USA}
\newcommand{\Granada}{Universidad de Granada, Granada E-18071, Spain}
\newcommand{\IIT}{Illinois Institute of Technology (IIT), Chicago, IL 60616, USA}
\newcommand{\ICL}{Imperial College London, London SW7 2AZ, United Kingdom}
\newcommand{\Indiana}{Indiana University, Bloomington, IN 47405, USA}
\newcommand{\Kansas}{The University of Kansas, Lawrence, KS, 66045, USA}
\newcommand{\KSU}{Kansas State University (KSU), Manhattan, KS, 66506, USA}
\newcommand{\Lancaster}{Lancaster University, Lancaster LA1 4YW, United Kingdom}
\newcommand{\LANL}{Los Alamos National Laboratory (LANL), Los Alamos, NM, 87545, USA}
\newcommand{\Louisiana}{Louisiana State University, Baton Rouge, LA, 70803, USA}
\newcommand{\Manchester}{The University of Manchester, Manchester M13 9PL, United Kingdom}
\newcommand{\MIT}{Massachusetts Institute of Technology (MIT), Cambridge, MA, 02139, USA}
\newcommand{\Michigan}{University of Michigan, Ann Arbor, MI, 48109, USA}
\newcommand{\MSU}{Michigan State University, East Lansing, MI 48824, USA}
\newcommand{\Minnesota}{University of Minnesota, Minneapolis, MN, 55455, USA}
\newcommand{\Nankai}{Nankai University, Nankai District, Tianjin 300071, China}
\newcommand{\NMSU}{New Mexico State University (NMSU), Las Cruces, NM, 88003, USA}
\newcommand{\Oxford}{University of Oxford, Oxford OX1 3RH, United Kingdom}
\newcommand{\Pitt}{University of Pittsburgh, Pittsburgh, PA, 15260, USA}
\newcommand{\QMUL}{Queen Mary University of London, London E1 4NS, United Kingdom}
\newcommand{\Rutgers}{Rutgers University, Piscataway, NJ, 08854, USA}
\newcommand{\SLAC}{SLAC National Accelerator Laboratory, Menlo Park, CA, 94025, USA}
\newcommand{\SDSMT}{South Dakota School of Mines and Technology (SDSMT), Rapid City, SD, 57701, USA}
\newcommand{\Maine}{University of Southern Maine, Portland, ME, 04104, USA}
\newcommand{\Syracuse}{Syracuse University, Syracuse, NY, 13244, USA}
\newcommand{\TelAviv}{Tel Aviv University, Tel Aviv, Israel, 69978}
\newcommand{\UTA}{University of Texas, Arlington, TX, 76019, USA}
\newcommand{\Tufts}{Tufts University, Medford, MA, 02155, USA}
\newcommand{\VTech}{Center for Neutrino Physics, Virginia Tech, Blacksburg, VA, 24061, USA}
\newcommand{\Warwick}{University of Warwick, Coventry CV4 7AL, United Kingdom}

\affiliation{\ANL}
\affiliation{\Bern}
\affiliation{\BNL}
\affiliation{\UCSB}
\affiliation{\Cambridge}
\affiliation{\CIEMAT}
\affiliation{\Chicago}
\affiliation{\Cincinnati}
\affiliation{\CSU}
\affiliation{\Columbia}
\affiliation{\Edinburgh}
\affiliation{\FNAL}
\affiliation{\Granada}
\affiliation{\IIT}
\affiliation{\ICL}
\affiliation{\Indiana}
\affiliation{\Kansas}
\affiliation{\KSU}
\affiliation{\Lancaster}
\affiliation{\LANL}
\affiliation{\Louisiana}
\affiliation{\Manchester}
\affiliation{\MIT}
\affiliation{\Michigan}
\affiliation{\MSU}
\affiliation{\Minnesota}
\affiliation{\Nankai}
\affiliation{\NMSU}
\affiliation{\Oxford}
\affiliation{\Pitt}
\affiliation{\QMUL}
\affiliation{\Rutgers}
\affiliation{\SLAC}
\affiliation{\SDSMT}
\affiliation{\Maine}
\affiliation{\Syracuse}
\affiliation{\TelAviv}
\affiliation{\UTA}
\affiliation{\Tufts}
\affiliation{\VTech}
\affiliation{\Warwick}

\author{P.~Abratenko} \affiliation{\Tufts}
\author{D.~Andrade~Aldana} \affiliation{\IIT}
\author{L.~Arellano} \affiliation{\Manchester}
\author{J.~Asaadi} \affiliation{\UTA}
\author{A.~Ashkenazi}\affiliation{\TelAviv}
\author{S.~Balasubramanian}\affiliation{\FNAL}
\author{B.~Baller} \affiliation{\FNAL}
\author{A.~Barnard} \affiliation{\Oxford}
\author{G.~Barr} \affiliation{\Oxford}
\author{D.~Barrow} \affiliation{\Oxford}
% ...

\collaboration{The MicroBooNE Collaboration}
\thanks{microboone\_info@fnal.gov}\noaffiliation

\date{\today}

\begin{abstract}
We report results from an updated search for neutral current (NC) resonant $\Delta$(1232) baryon production and subsequent $\Delta$ radiative decay (NC $\Delta\rightarrow N \gamma$). We consider events with and without final state protons; events with a proton can be compared with the kinematics of a $\Delta(1232)$ baryon decay, while events without a visible proton represent a more generic phase space. In order to maximize sensitivity to each topology, we simultaneously make use of two different reconstruction paradigms, Pandora and Wire-Cell, which have complementary strengths, and select mostly orthogonal sets of events. 
Considering an overall scaling of the NC $\Delta\rightarrow N \gamma$ rate as an explanation of the MiniBooNE anomaly, our data exclude this hypothesis at 94.4\% CL. 
When we decouple the expected correlations between NC $\Delta \rightarrow N
\gamma$ events with and without final state protons, our data exclude an interpretation in which all excess events have associated protons at 2.0$\sigma$, and are consistent with an interpretation in which all excess events have no associated protons at 0.63$\sigma$.

\end{abstract}

\maketitle

\italicheading{Introduction.} The 4.8$\sigma$ MiniBooNE low-energy excess (LEE) of electron-like neutrino interactions \cite{miniboone_lee_latest} remains an important unexplained result in particle physics \cite{nf02_whitepaper}. 
There have been many attempts to explain this excess 
as additional electrons, photons, or electron-positron pairs, produced by
standard-model (SM) or beyond-the-standard-model (BSM) hypotheses \cite{heavy_sterile_beam, 250_MeV_sterile_decay, heavy_neutrino_radiative, mixed_oscillation_decay, dark_neutrino_portal, dark_seesaw, heavy_sterile, two_higgs_doublet, extended_higgs, axion_like, 17MeV_pseudoscalar}.
As a Cherenkov detector, MiniBooNE was largely unable to differentiate these different hypotheses, and therefore each possibility must be investigated. In contrast, the MicroBooNE liquid argon time projection chamber \cite{uboone_detector} has high-resolution 3D imaging and calorimetry, allowing for excellent electron-photon separation. MicroBooNE operated in the same Booster Neutrino Beam (BNB) at approximately the same baseline as MiniBooNE, giving it the capability to investigate the LEE in detail.

In this Letter, we present an updated test of a single-photon interpretation of the MiniBooNE LEE. This builds on a previous result \cite{glee_prl} that searched for neutrino-induced neutral current $\Delta$ radiative decay to a nucleon and a photon (\ncdeltarad); an anomalous enhancement of this interaction rate by a factor of 3.18, which could explain the MiniBooNE LEE \cite{miniboone_lee_latest}, was disfavored. 
The previous result had significant sensitivity to events containing just one visible photon and one visible proton (\onegOnep), but limited sensitivity to events containing just one visible photon and zero visible protons (\onegZerop). 
We expand the previous result by incorporating similar selections using different reconstruction tools, targeting a broader signal category with enhanced sensitivity to the signal hypothesis.
The analysis presented in this letter features significantly enhanced sensitivity to \onegZerop\ events.

Although the PDG \cite{pdg_2024} assigns only an 8.3\% uncertainty to the $\Delta\rightarrow N\gamma$ branching fraction, an enhancement due to unmodeled nuclear effects or new physics on the NC $\Delta\rightarrow N\gamma$ rate remains an interesting model for potential single-photon excesses.
This process has never been observed in neutrino scattering, and it is the only significant expected source of single photons in MiniBooNE and MicroBooNE. Thus, it is a natural process to consider when trying to connect observations between the two detectors. The \ncdeltarad\ process allows for a comparison of single-photon event rates between MiniBooNE and MicroBooNE, accounting for beam exposure, nuclear modeling, and selection efficiencies. 
Additionally, a scaling of \ncdeltarad\ events is the only quantitative measure of a single-photon excess reported by the MiniBooNE collaboration \cite{miniboone_lee_latest}, allowing for a direct comparison between MicroBooNE and MiniBooNE photon observations. A search for NC $\Delta\rightarrow N\gamma$ events can also be sensitive to other types of neutrino-induced neutral current single-photon production \cite{nc_gamma_intermediate_energies}.

\italicheading{Selections.} We use the same selections as Ref.~\cite{glee_prl} using Pandora \cite{pandora} reconstruction, and we add new selections developed using Wire-Cell (WC) \cite{wc_imaging_matching} reconstruction. 
Pandora and Wire-Cell are complementary approaches to event reconstruction, with Pandora performing provisional clustering of 2D hits in each wire plane before correlating features across planes to produce 3D particles, while Wire-Cell uses a tomographic approach to first correlate 2D hits across planes before proceeding with 3D pattern recognition to produce 3D particles.
In each case, selections were developed in order to maximize the number of signal \ncdeltarad\ events while minimizing all other backgrounds. The Pandora selections which are unchanged relative to Ref.~\cite{glee_prl}, use preselections targeting a specific topology, \onegOnep\ or \onegZerop, and then use ensembles of Boosted Decision Trees (BDTs) targeting different background types. The Wire-Cell selections use a generic neutrino preselection \cite{wc_generic} followed by a single BDT trained to select \ncdeltarad\ events from all topologies. The Wire-Cell BDT is trained on a large number of reconstructed variables, in a similar method as the charged-current (CC) $\nu_e$ BDT in Ref.~\cite{wc_elee_prd}. 
After applying the Wire-Cell BDT requirement, we split the selection into \onegNp\ and \onegZerop\ using a 35 MeV reconstructed proton kinetic energy threshold. This choice is comparable to the corresponding effective threshold in Pandora proton track reconstruction, and corresponds to a proton that travels about 1 centimeter, a few wire spacings, the minimum range necessary to perform particle identification using reconstructed $dE/dx$ measurements. Unlike the Pandora selections, which contain only events with zero or one reconstructed proton and zero reconstructed charged pions $1\gamma 0p 0\pi^\pm + 1\gamma 1p 0\pi^\pm$, the Wire-Cell selections do not reject events with two or more reconstructed protons or events with one or more reconstructed charged pions in the final state, making the reconstructed topology $1\gamma Xp X\pi^\pm$, where $X$ refers to any number of particles. These relaxed particle multiplicity requirements increase the relative \ncdeltarad\ selection efficiency over a combined $1\gamma 0p 0\pi^\pm$ and $1\gamma 1p 0\pi^\pm$ Wire-Cell selection by 9\%, and could increase sensitivity to more complex single-photon hypotheses, for example those involving two nucleons as described in Ref.~\cite{2p2hgamma}.

We investigate events with ($Np$) and without ($0p$) reconstructed protons separately because these selections can point towards different types of physics effects. \ncdeltarad\ events with no hadronic activity represent a phase space with only 2 degrees of freedom, shower energy, and shower angle. Therefore, our \onegZerop\ selection is not as sensitive to the underlying physical source of the photon as our \onegNp\ selection, which preferentially selects events with photon-proton invariant mass near the $\Delta$ resonance. Because of this, the \onegZerop\ channel can be tied to a broader set of alternative excess hypotheses, whether from SM backgrounds or BSM signatures.

Each selection results in a single-bin sample in reconstructed shower energy. The bins are 0-600, 100-700, and 0-1500 MeV for the Pandora \onegOnep, Pandora \onegZerop, and both Wire-Cell selections (\onegNp\ and \onegZerop), respectively. All selections were developed according to a blinding policy, where only a small sample of data corresponding to $5\times 10^{19}$ protons-on-target (POT) was examined before the selections were finalized. The Pandora and Wire-Cell samples used for the reported results correspond to $6.80\times 10^{20}$ and $6.37\times 10^{20}$ POT, respectively, due to different data processing campaigns.

Table~\ref{tab:effs_and_purs} shows a summary of the efficiency and purity of each selection. The purity in each selection is limited by events containing two photons from a $\pi^0$ decay in which only one photon was reconstructed. In particular, note the improvement in the Wire-Cell $1\gamma 0 p$ channel relative to the Pandora $1\gamma 0 p$ channel, and the large increase in total efficiency when all selections are combined.

\newcolumntype{C}[1]{>{\hspace{#1}}c<{\hspace{#1}}}

\begin{table}[H]
\centering
\caption{Efficiency and purity summary. The rightmost column shows the efficiency and purity for a union of all four selections; note that the combined efficiency is less than the sum of the four efficiencies, because some events can be selected by both reconstructions. Efficiency is calculated as the fraction of simulated true \ncdeltarad\ events in the fiducial volume which enter the final selection. Purity is calculated as the fraction of the predicted selected events which are from the \ncdeltarad\ process.}
\resizebox{0.49\textwidth}{!}{
\begin{tabular}{c C{0.2cm} c c c c c c} 
    \hline
    \hline
    & &\makecell{WC \\ $1\gamma Np$} & \makecell{Pandora \\ $1\gamma 1p$} & \makecell{WC \\ $1\gamma 0p$} & \makecell{Pandora \\ $1\gamma 0p$} & & \makecell{Combined} \\ 
    \hline
    \makecell{\ncdeltarad\ efficiency} & & 4.09\% & 4.24\% & 8.79\% & 5.52\% & &19.64\%\\
    \makecell{\ncdeltarad\ purity} & & 9.60\% & 14.84\% & 7.50\% & 3.98\% & &6.37\%\\
    \hline
    \hline
\end{tabular}
}
\label{tab:effs_and_purs}
\end{table}

Signal and background predictions for each selection are generated with Monte Carlo simulations. 
These model the neutrino flux, neutrino-argon interactions, and detector response. 
The simulated detector response is overlaid on cosmic ray backgrounds measured \textit{in-situ} with dedicated samples collected without the neutrino beam.
Simulated data samples were reprocessed for this analysis, leading to some differences between this work and the result reported in Ref.~\cite{glee_prl}; these differences fall within the statistical uncertainties of the simulated data sample. 

It is worth noting that the Pandora and Wire-Cell selections are almost orthogonal. Of the 175.6 predicted events in the Wire-Cell selection, only 21.9 are found in the 194.4-event Pandora selection. This small rate of overlap indicates that there is significant room for future improvements in single-photon reconstruction and selection; it also highlights the benefit from this analysis which combines the selected events from these two independent workflows.

\italicheading{Uncertainties.} We determine systematic uncertainties by following the same procedure as outlined in Ref.~\cite{wc_elee_prd}.

(1) We consider BNB flux uncertainties by varying $\pi^\pm$, $K^\pm$, and $K_L^0$ production rates, altering the beam line configuration modeling within its uncertainties, and fluctuating the pion and nucleon total, inelastic, and quasi-elastic scattering cross sections on beryllium and aluminum \cite{bnb_flux_uboone}.

(2) Neutrino-argon interaction cross section uncertainties are modeled using GENIE v3.0.6 tune G18\_10a\_02\_11a, (``MicroBooNE tune"), varying 46 underlying model parameters, including those related to the quasi-elastic, meson-exchange-current, resonance, deep-inelastic-scattering, coherent scattering, neutral current, and final state interaction models \cite{genie_v3, genie-tune-paper}. No GENIE \ncdeltarad\ branching ratio uncertainty was considered, as this was a free parameter in this analysis; this matches the systematic uncertainty treatment in Ref.~\cite{glee_prl}.

(3) Uncertainties on hadron-argon interactions outside of the struck nucleus are modeled by considering inelastic collisions of protons, positive pions, and negative pions with argon, varying each cross section around its mean \textsc{Geant4} prediction \cite{GEANT4:2002zbu} by 20\%.

(4) We consider detector uncertainties related to the electronic response to ionization charge, light yield, and propagation, space charge effect, and recombination model \cite{det_unc}.

(5) We consider Monte Carlo statistical uncertainties. Statistical uncertainty correlations from events selected by both Wire-Cell and Pandora are accounted for by a repeated sampling bootstrapping procedure \cite{wc_elee_prd}.

(6) We add an additional 50\% uncertainty for events with a true neutrino vertex outside the cryostat in order to consider any possible mismodeling of external materials.

Additional systematic uncertainties associated with higher mass resonance radiative decays, photonuclear absorption, and coherent single-photon production are negligible in this analysis. The relative sizes of all uncertainties on our signal channels are shown in Table~\ref{tab:sys_breakdown}. 

\begin{table}[tb]
\footnotesize
\centering
\caption{Signal channel systematic uncertainty breakdown.}
\begin{tabular}{c c c c c} 
    \hline
    \hline
    Uncertainty Type & \makecell{WC\\$1\gamma N p$} & \makecell{WC\\$1\gamma0p$} & \makecell{Pandora\\$1\gamma 1 p$} & \makecell{Pandora\\$1\gamma0p$} \\
    \hline
    Flux model & 6.58\% & 6.29\% & 7.39\% & 6.66\% \\
    GENIE cross section & 19.49\% & 17.09\% & 25.96\% & 17.87\% \\
    Hadron re-interaction & 1.27\% & 0.70\% & 2.22\% & 0.89\% \\
    Detector modeling & 17.58\% & 23.35\% & 15.69\% & 10.96\% \\
    \makecell{Monte Carlo statistics} & 5.64\% & 3.67\% & 10.40\% & 5.47\% \\
    \makecell{Out-of-cryostat interactions} & 0.00\% & 0.33\% & 0.00\% & 1.02\% \\
    \noalign{\vskip 7 pt}
    \makecell{Total uncertainty\\(unconstrained)} & 27.65\% & 29.85\% & 32.94\% & 22.61\% \\
    \noalign{\smallskip}
    \makecell{Total uncertainty\\(constrained)} & 16.80\% & 12.39\% & 23.96\% & 15.02\% \\
    \hline
    \hline
\end{tabular}
\label{tab:sys_breakdown}
\end{table}

In order to reduce systematic uncertainties and adjust the central-value prediction in a data-driven way, we apply a conditional constraint based on the measurement of \ncpio\ and $\nu_\mu$CC events from dedicated sidebands.
This constraint considers all statistical and systematic uncertainties and follows the same procedure as the constraint applied in Ref.~\cite{wc_elee_prd}. We use the same sideband channels to constrain all four signal channels. The constraining \ncpio\ selections use Wire-Cell reconstruction, and are updated relative to the \ncpio\ selection in Ref.~\cite{wc_elee_prd} by utilizing a BDT, described in Ref.~\cite{microboone_2d_nc_pi0}. The constraining $\nu_\mu$CC selections also use Wire-Cell reconstruction, and are identical to those in Ref.~\cite{wc_elee_prd}. As shown in Table \ref{tab:components}, 
the largest background contribution to our signal channels comes from \ncpio\ interactions, and this component is significantly constrained by the observation in the \ncpio\ selections. The \ncpio\ selections also constrain the signal \ncdeltarad\ events, which have large correlations with many \ncpio\ interactions because of the common $\Delta$ resonance parentage. The $\nu_\mu$CC selections further constrain some uncertainties. These constraining channels are split into reconstructed energy distributions with and without reconstructed protons, which can be found in the Supplemental Material \cite{supplemental_material}. 

\begin{figure}[tb]
    \centering
    \includegraphics[trim=35 50 70 80, clip, width=0.49\textwidth]{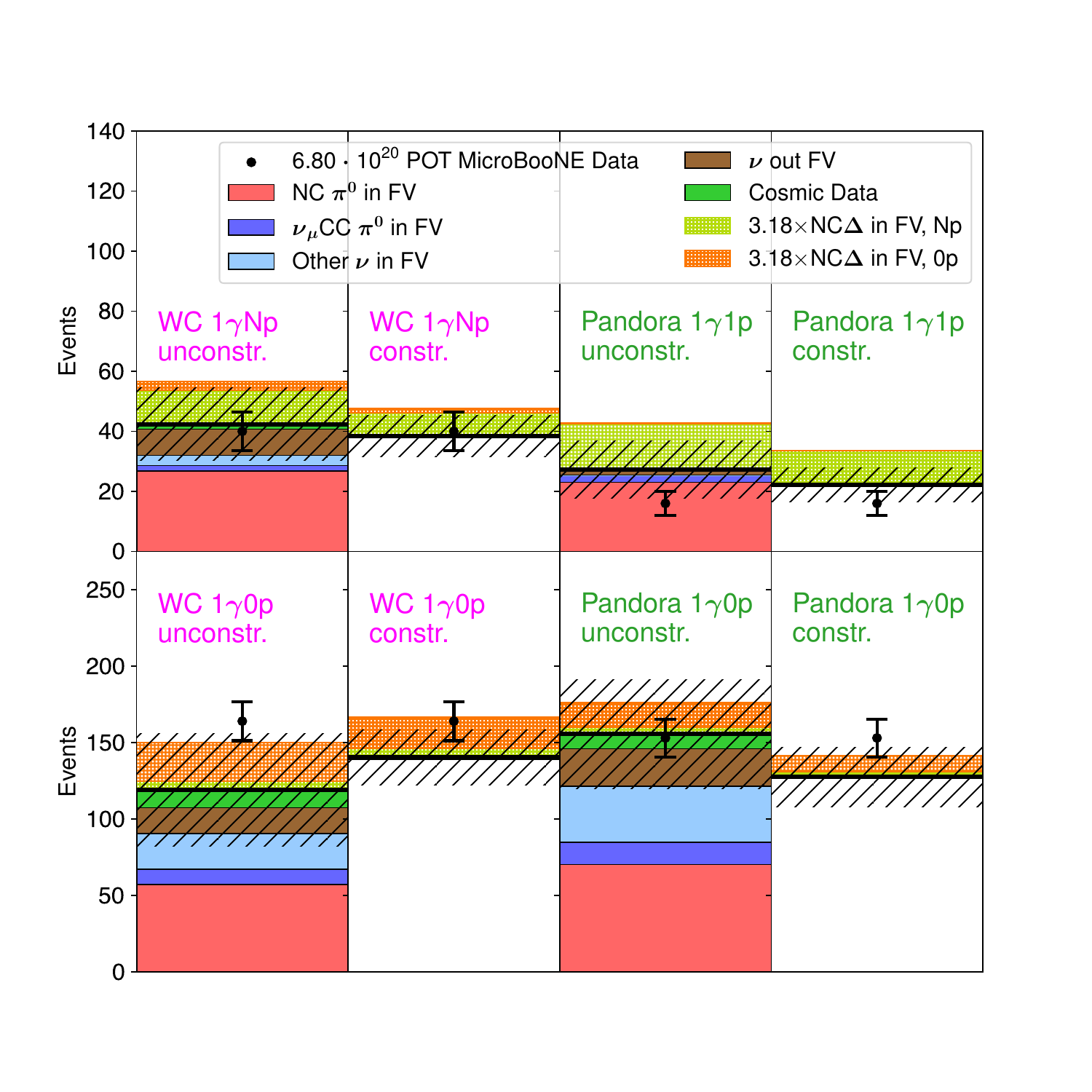}
    \caption{Wire-Cell and Pandora signal channels, unconstrained and constrained. The no-\ncdeltarad\ prediction is shown in black, with diagonal hashes indicating the systematic uncertainty. The LEE prediction with a $x_\Delta=3.18$ enhancement of the nominal \ncdeltarad\ is shown in green and orange for signal with and without true final state protons with kinetic energy of at least 35 MeV, respectively. The Pandora and Wire-Cell samples correspond to $6.80\times 10^{20}$ and $6.37\times 10^{20}$ POT, respectively.}
    \label{fig:signals}
\end{figure}

\begin{figure}[tb]
    \centering
    \includegraphics[width=0.49\textwidth]{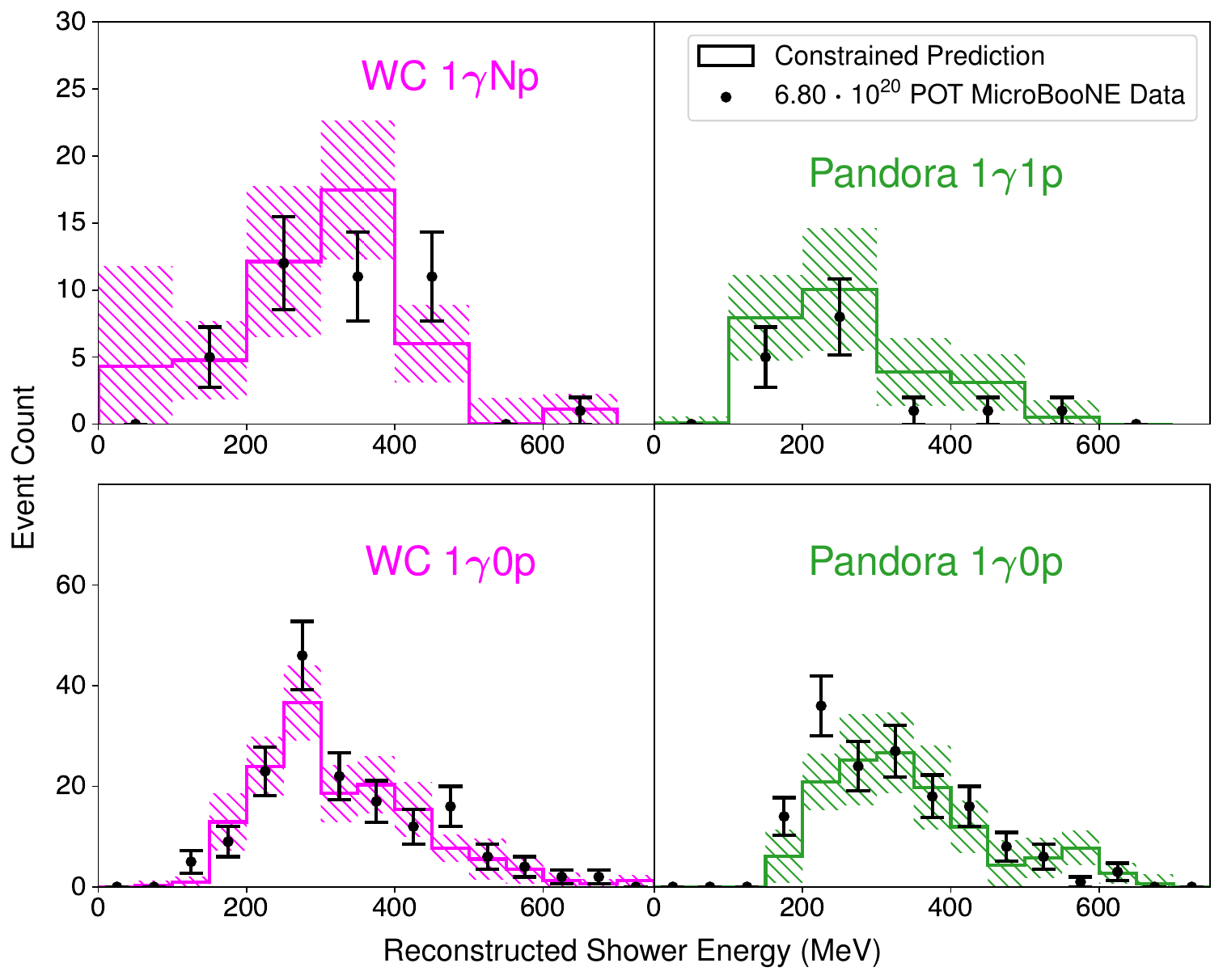}
    \caption{Wire-Cell and Pandora signal channel shower energy distributions, constrained by sideband observations. The prediction shows the nominal \ncdeltarad\ scaling, $x_\Delta=1$. The top panels have bin widths of 100 MeV, while the bottom panels have bin widths of 50 MeV. In each panel, the rightmost bin is an overflow bin. The Pandora and Wire-Cell samples correspond to $6.80\times 10^{20}$ and $6.37\times 10^{20}$ POT, respectively.}
    \label{fig:energy_constr}
\end{figure}

\italicheading{Results.} Our four signal channels are shown with and without the conditional constraint in Table~\ref{tab:components} and Fig.~\ref{fig:signals}, and the resulting constrained shower energy distributions are shown in Fig.~\ref{fig:energy_constr}. The signal channel uncertainties before and after constraint are shown in Table~\ref{tab:sys_breakdown}. The constraints generally act to lower the prediction, due to an observed overprediction of \ncpio\ events containing at least one proton. However, the observed underprediction of low energy \ncpio\ events with no protons acts to increase the prediction for the Wire-Cell \onegZerop\ channel. Wire-Cell selected events with multiple protons and selected events with charged pions each agree with our nominal predictions within uncertainties.

\begin{table}[tb]
\footnotesize
\centering
\caption{Signal and background components. Categories are broken into those with true neutrino interaction vertices inside and outside the fiducial volume (FV). The signal is denoted by \ncdeltarad\ in FV.}
\begin{tabular}{c c c c c} 
    \hline
    \hline
    Process & \makecell{WC\\$1\gamma N p$} & \makecell{WC\\$1\gamma0p$} & \makecell{Pandora\\$1\gamma 1 p$} & \makecell{Pandora\\$1\gamma0p$} \\
    \hline
    NC $1\pi^0$ in FV & 26.8 & 57.2 & 23.0 & 70.1 \\
    CC $1\pi^0$ in FV & 1.9 & 10.0 & 2.4 & 14.7 \\
    Other $\nu$ in FV & 8.7 & 16.9 & 1.9 & 24.6 \\
    Out FV & 3.4 & 23.3 & 0.0 & 36.6 \\
    Cosmic Beam-off Data & 1.6 & 11.7 & 0.0 & 9.8 \\
    \noalign{\vskip 7 pt}
    NC $\Delta\rightarrow N \gamma$ in FV & 4.5 & 9.7 & 4.9 & 6.5 \\
    \noalign{\vskip 7 pt}
    \makecell{Unconstrained\\total prediction} & 46.8 & 128.7 & 32.2 & 162.2 \\
    \noalign{\smallskip}
    \makecell{Constrained\\total prediction} & 41.3 & 148.5 & 25.8 & 131.9 \\
    \noalign{\smallskip}
    \makecell{Observed data} & 40 & 164 & 16 & 153 \\
    \hline
    \hline
\end{tabular}
\label{tab:components}
\end{table}

We consider two types of MiniBooNE LEE hypotheses. 
We first consider a simple scaling where we vary the total \ncdeltarad\ cross section equally across all samples. This is the same procedure employed in MicroBooNE’s previous \ncdeltarad\ search \cite{glee_prl}.
In this analysis, we also consider a second scaling that allows for the possibility of different rates of \ncdeltarad\ for the final states with and without protons. In this model, the rates of these two subprocesses are allowed to vary independently, leading to a model with 2 degrees of freedom. 

\begin{figure}[tb]
    \centering
    \includegraphics[trim=10 20 10 20, clip, width=0.49\textwidth]{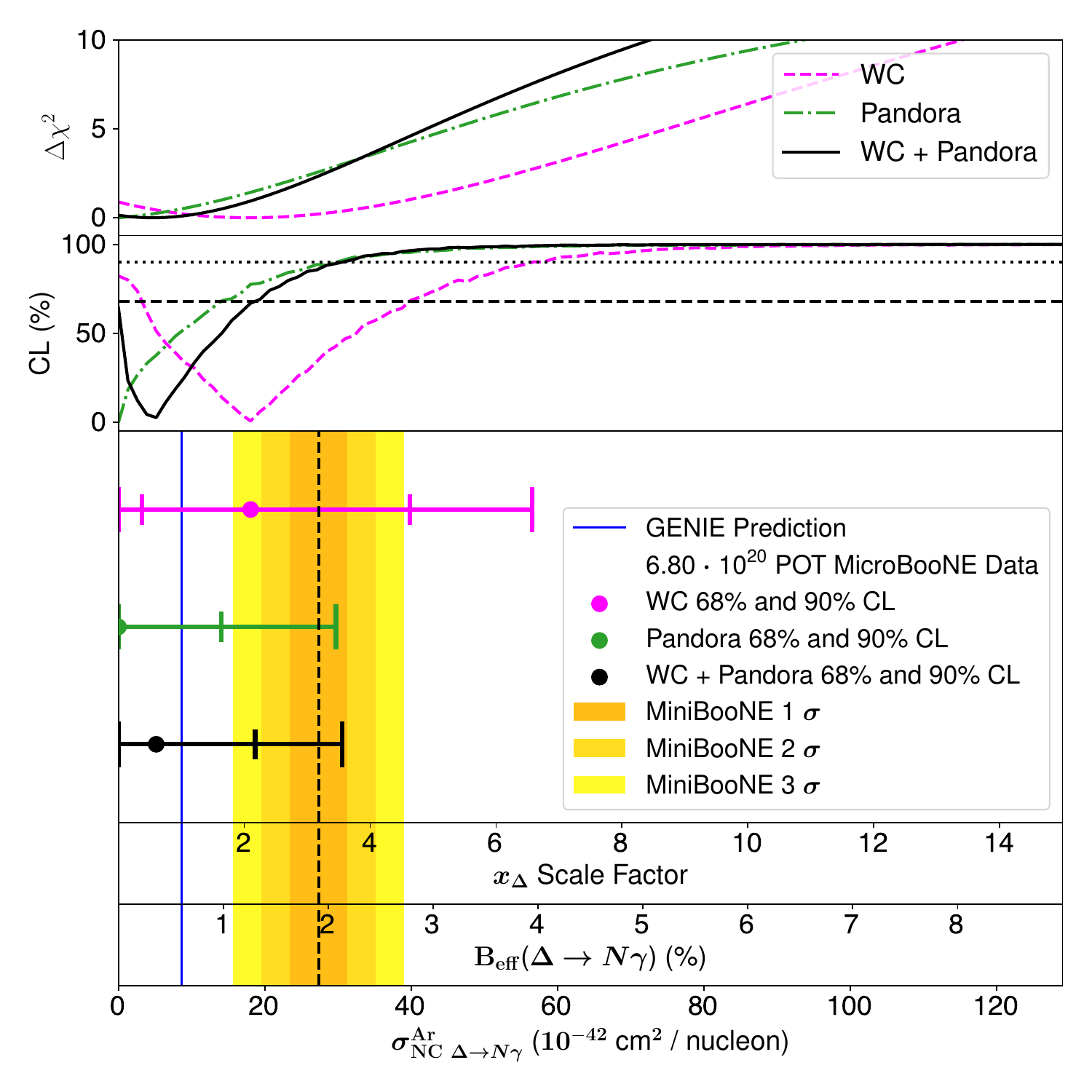}
    \caption{\ncdeltarad\ scaling exclusions. Black horizontal dashed lines indicate 68\% and 90\% CL values. The effective branching fraction and cross section are simple rescalings of the $x_\Delta$ scale factor. The Pandora and Wire-Cell samples correspond to $6.80\times 10^{20}$ and $6.37\times 10^{20}$ POT, respectively.}
    \label{fig:one_d}
\end{figure}

For the One-dimensional LEE hypothesis, we fit the signal and constraining channels with a single free parameter $x_\Delta$, corresponding to the normalization of the nominal rate of \ncdeltarad\ events. A value of one corresponds to the standard GENIE prediction, and a value of 3.18 corresponds to the MiniBooNE LEE under a \ncdeltarad\ scaling hypothesis \cite{miniboone_lee_latest}.
To compare with MiniBooNE visually in Fig.~\ref{fig:one_d}, we assign a $1\sigma$ confidence interval for the scaling parameter of $3.18\pm0.45$, which has been estimated from the $4.8\sigma$ significance of the MiniBooNE LEE.
The $x_\Delta$ scaling parameter is also interpreted as a scaling of the effective branching fraction $B_\text{eff}(\Delta\rightarrow N \gamma)$ and as a scaling of the flux-averaged cross section for \ncdeltarad\ interactions on argon $\sigma^\text{Ar}_{\text{NC}\Delta\rightarrow N \gamma}$.

We form confidence intervals using the Feldman-Cousins approach \cite{feldman_cousins}. We use a Combined-Neyman-Pearson $\chi^2$ \cite{cnp_paper} and use a covariance matrix that includes systematic uncertainties and correlations between our four one-bin signal channels and all of our constraining bins.
This test is essentially the same one performed in Ref.~\cite{glee_prl}, with different signal channels and constraining channels, and small differences in the systematic uncertainty treatment. 
With the combination of Wire-Cell and Pandora selections, our expected 90\% CL upper limit exclusion is at $x_\Delta=3.18$, indicating notably higher sensitivity than either Pandora alone at $x_\Delta=4.00$, or Wire-Cell alone at $x_\Delta=4.15$. More details can be found in the Supplemental Material.
The result is shown in Fig.~\ref{fig:one_d}. We see consistency with both the standard GENIE prediction and with the MiniBooNE LEE under an $x_\Delta=3.18$ hypothesis within 90\% CL. This is the case for all three sets of data considered: Wire-Cell, Pandora, and Wire-Cell + Pandora. 
The Wire-Cell + Pandora result has a best fit that lies slightly below the GENIE prediction, includes $x_\Delta=0$ at 68\% CL, and includes $x_\Delta=3.18$ at 90\% CL.
The Pandora selections prefer lower-scale factors, while the Wire-Cell selections prefer higher-scale factors. The result for the Pandora-only exclusion is consistent with the result in Ref.~\cite{glee_prl}, but these are not identical due to the different sideband constraints in this work.

We also perform a two-hypothesis test, using a $\Delta\chi^2$ test statistic comparing the MiniBooNE LEE under an $x_\Delta=3.18$ hypothesis and the standard GENIE prediction. We exclude the LEE hypothesis with a p-value of 94.4\% CL. When simply comparing the nominal and LEE predictions, this two-hypothesis test is more sensitive than the one-dimensional scaling test described above, and therefore we rely on it for our main conclusion.

With a two-dimensional LEE hypothesis, we can consider each final state separately, and decouple the search for an excess in \ncdeltarad\ events from the predicted breakdown of hadronic activity as modeled in the GENIE neutrino interaction generator. 
We test this quantitatively by considering separate scalings of signal \ncdeltarad\ events with and without true primary protons with kinetic energy greater than 35 MeV. We call these scaling parameters $x_{Np}$ and $x_{0p}$, respectively.

In order to translate the inclusive \ncdeltarad\ excess at each point in the $(x_{Np}, x_{0p})$ space, we split signal events according to the formula $0.53 \cdot x_{Np} + 0.47\cdot x_{0p}$, based on our modeling of the make up of $0p$ and $Np$ signatures for signal events in MicroBooNE. We then estimate the significance at each point in this 2D space using the same method as for the 1D fit.
Note that we do not make any assumptions about true proton multiplicities for \ncdeltarad\ events in MiniBooNE and instead only consider the total predicted count. 

\begin{figure*}[t]
  \centering
  \begin{minipage}[t]{0.49\textwidth}
    \centering
    \includegraphics[trim=60 50 95 110, clip, width=\textwidth]{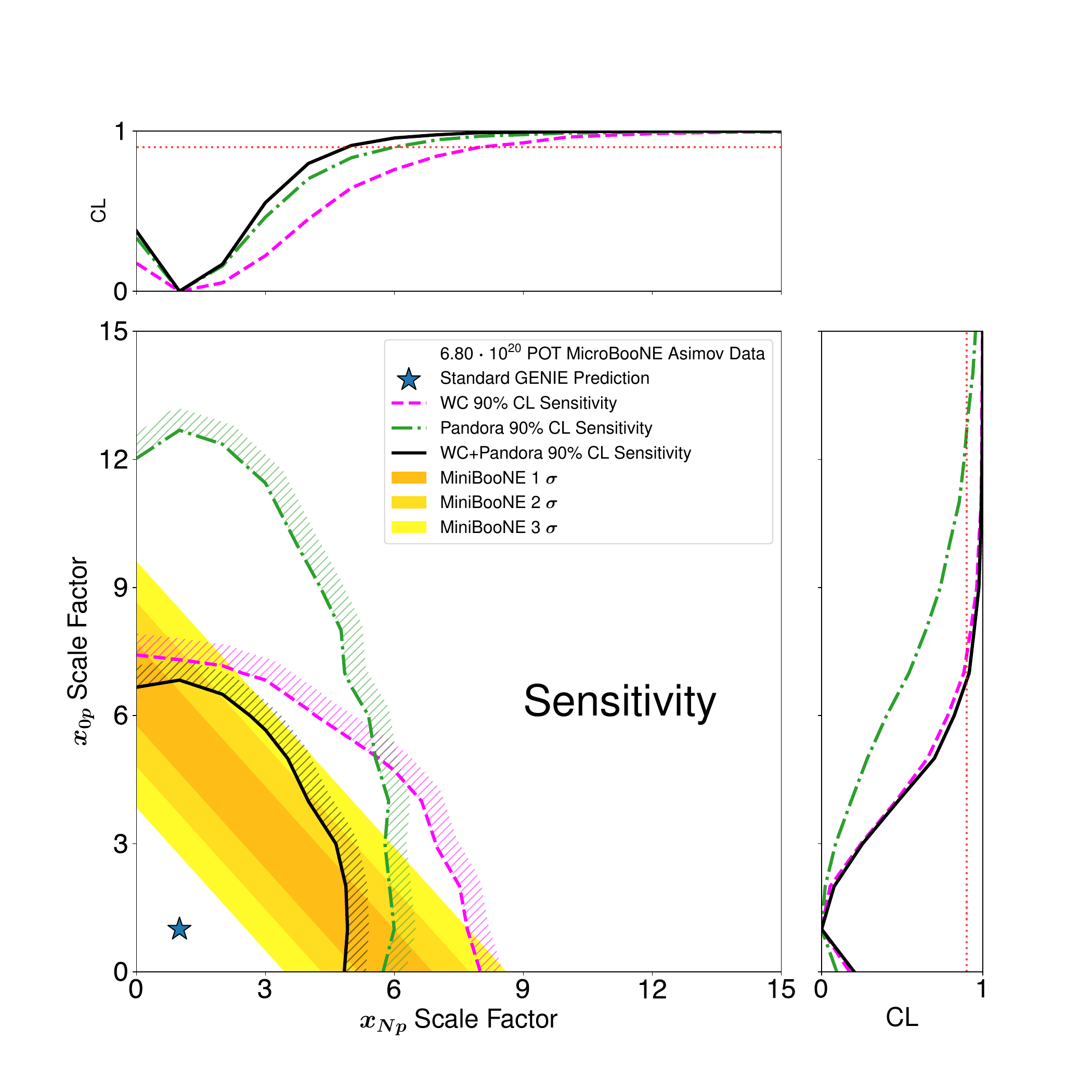}
    \caption{Two-dimensional $x_{\Delta Np}$ and $x_{\Delta 0p}$ scaling exclusion sensitivity with Asimov data, a fake data set that exactly matches the prediction. One-dimensional profiles are shown in the top and right, with a dashed line indicating 90\% CL. The hashed region indicates the side of each curve which is being excluded. The Pandora and Wire-Cell Asimov data samples correspond to $6.80\times 10^{20}$ and $6.37\times 10^{20}$ POT, respectively.}
    \label{fig:two_d_asimov}
  \end{minipage}\hfill
  \begin{minipage}[t]{0.49\textwidth}
    \centering
    \includegraphics[trim=60 50 95 110, clip, width=\textwidth]{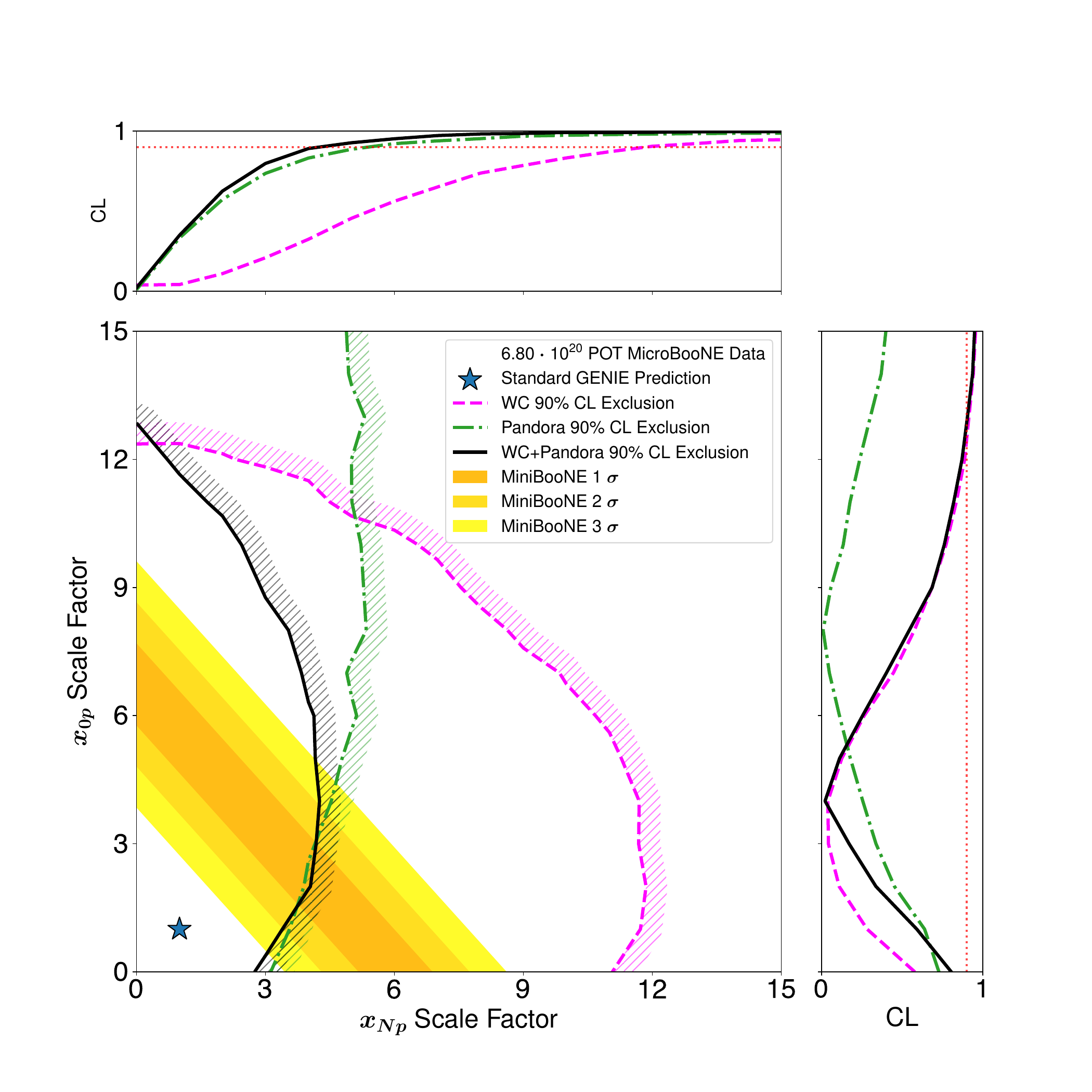}
    \caption{Two-dimensional $x_{Np}$ and $x_{0p}$ scaling data exclusions. One-dimensional profiles are shown in the top and right, with a dashed line indicating 90\% CL. The hashed region indicates the side of each curve which is being excluded. The Pandora and Wire-Cell data samples correspond to $6.80\times 10^{20}$ and $6.37\times 10^{20}$ POT, respectively.}
    \label{fig:two_d}
  \end{minipage}
\end{figure*}

We apply a Feldman-Cousins procedure, the same as was used to obtain the results in Fig.~\ref{fig:one_d}, on a two-dimensional space of hypotheses to extract the exclusion contours. The expected sensitivities are shown in Fig.~\ref{fig:two_d_asimov}, while the exclusions using real data are shown in Fig.~\ref{fig:two_d}. The Wire-Cell-only contour in pink excludes large \onegNp\ and large \onegZerop\ scalings about equally well. The Pandora-only contour in green excludes \onegNp\ scalings well, but provides a weaker constraint on \onegZerop\ scalings. This is expected due to the slight overprediction in the Pandora \onegOnep\ channel and the weak sensitivity of the Pandora \onegZerop\ channel. The Wire-Cell+Pandora combined result disfavors higher scaling values for true \onegNp\ events, but does not exclude higher scaling values for true \onegZerop\ events. This behavior is explained by the overprediction in the Pandora \onegOnep\ channel, and the underprediction in the Wire-Cell and Pandora \onegZerop\ channels. Due to the weaker correlations between \onegNp\ and \onegZerop\ signal predictions, this two-dimensional test leads to weaker exclusions than the one-dimensional test. The resulting exclusion and the sensitivity are stronger for the combined Wire-Cell+Pandora result than the exclusions with either reconstruction alone.

In simple scalings of the \ncdeltarad\ rate, the data are found to be consistent with the nominal prediction and disfavor the \ncdeltarad\ scaling LEE prediction. Meanwhile, with a more general LEE model which considers different scalings for $0p$ and $Np$ events, our data are consistent with the nominal prediction and exclude \ncdeltarad-like explanations of the MiniBooNE LEE where all single-photon events are assumed to have associated proton activity. 
Focusing on two specific points in this phase space, our data exclude the $(x_{Np}, x_{0p}) = (5, 1)$ point, an interpretation in which all excess events have associated protons, at 2.0$\sigma$, and our data are consistent with the $(x_{Np}, x_{0p}) = (1, 6)$ point, an interpretation in which all excess events have no associated protons, at 0.63$\sigma$.
The majority of the LEE exclusion power comes from the Pandora $1\gamma 1 p$ channel with its data deficit. However, the Wire-Cell channels increase the sensitivity and exclusion power, most notably for events with no visible protons.

\italicheading{Conclusions.} Our updated search for NC resonant $\Delta(1232)$ production and subsequent radiative decay, utilizing both the Pandora and Wire-Cell reconstruction techniques, yields significant constraints on interpretations of the MiniBooNE LEE. Under the assumption of a uniform scaling of the \ncdeltarad\ rate, our analysis excludes this hypothesis at 94.4\% CL, consistent with our previous result \cite{glee_prl}. Furthermore, when considering a model that permits independent scaling for events with and without final state protons, our results rule out scenarios where the majority of the excess events are associated with protons, while remaining compatible with cases where most excess events occur without a visible proton. 
MicroBooNE has also investigated other types of single photons, including NC coherent single-photon production \cite{uboone_coherent_gamma} and an inclusive search for single photons \cite{uboone_inclusive_gamma}. The analysis presented here uses approximately half of MicroBooNE's collected BNB data set, and future analyses will use increased statistics, improved reconstructions, and different signal models to further advance our understanding of single-photon events in MicroBooNE. A data release corresponding to this analysis is available at Ref. \cite{nc_delta_hepdata}.

\italicheading{Acknowledgments.} We acknowledge the contributions of technical and scientific staff 
to the design, construction, and operation of the MicroBooNE detector 
as well as the contributions of past collaborators to the development 
of MicroBooNE analyses, without whom this work would not have been 
possible. 
This document was prepared by the MicroBooNE collaboration using the
resources of the Fermi National Accelerator Laboratory (Fermilab), a
U.S. Department of Energy, Office of Science, Office of High Energy Physics (HEP) User Facility.
Fermilab is managed by Fermi Forward Discovery Group, LLC, acting
under Contract No. 89243024CSC000002.  MicroBooNE is supported by the
following: 
the U.S. Department of Energy, Office of Science, Offices of High Energy Physics and Nuclear Physics; 
the U.S. National Science Foundation; 
the Swiss National Science Foundation; 
the Science and Technology Facilities Council (STFC), part of the United Kingdom Research and Innovation; 
the Royal Society (United Kingdom); 
the UK Research and Innovation (UKRI) Future Leaders Fellowship; 
and the NSF AI Institute for Artificial Intelligence and Fundamental Interactions. 
Additional support for 
the laser calibration system and cosmic ray tagger was provided by the 
Albert Einstein Center for Fundamental Physics, Bern, Switzerland. 
For the purpose of open access, the authors have applied 
a Creative Commons Attribution (CC BY) public copyright license to 
any Author Accepted Manuscript version arising from this submission.

%


\end{document}